# Foundations of consistent couple stress theory

Ali R. Hadjesfandiari, Gary F. Dargush

*Department of Mechanical and Aerospace Engineering*
*University at Buffalo, State University of New York*
*Buffalo, NY 14260 USA*

ah@buffalo.edu,   gdargush@buffalo.edu

July 29, 2015

**Abstract**

In this paper, we examine the recently developed skew-symmetric couple stress theory and demonstrate its inner consistency, natural simplicity and fundamental connection to classical mechanics. This hopefully will help the scientific community to overcome any ambiguity and skepticism about this theory, especially the validity of the skew-symmetric character of the couple-stress tensor. We demonstrate that in a consistent continuum mechanics, the response of infinitesimal elements of matter at each point decomposes naturally into a rigid body portion, plus the relative translation and rotation of these elements at adjacent points of the continuum. This relative translation and rotation captures the deformation in terms of stretches and curvatures, respectively. As a result, the continuous displacement field and its corresponding rotation field are the primary variables, which remarkably is in complete alignment with rigid body mechanics, thus providing a unifying basis. For further clarification, we also examine the deviatoric symmetric couple stress theory that, in turn, provides more insight on the fundamental aspects of consistent continuum mechanics.

**1. Introduction**

From the middle of the twentieth century onwards, there has been a shift towards developing continuum mechanics primarily from a thermodynamics perspective. As a result, much progress has been made, especially in constitutive modeling. However, this change in direction also has led to a departure of the discipline from the foundations of mechanics in its classical form, in which the fundamental entities are forces and couples, along with their kinematic conjugate displacements and rotations, respectively. Of course, the former relate directly to the basic conservation laws of linear and angular momentum, while the latter describe the pure rigid body motion. In rigid body mechanics, the force and moment equations are the governing equations describing the translational and rotational motion of the body in space. Consequently, it seems in developing a consistent continuum mechanics theory, we need to consider the rigid body portion of motion of infinitesimal elements of matter at each point of the continuum. This requires the inclusion of force- and couple-stresses in the formulation. Since the displacements and rotations at each point are the degrees of freedom of the infinitesimal body, the fundamental mechanical equations are still the force and moment equations at each point. However, to have a complete set of equations, we need the constitutive equations. This in turn requires consideration of the deformation or, more specifically, the relative rigid body motion of infinitesimal elements of matter at adjacent points of the continuum.

Cauchy elasticity, as the first continuum theory, focused on force-stresses and displacements. Couple-stresses were simply dismissed from the very beginning and, as a result, the moment equations merely provide the symmetric character of the force-stress tensor. Consequently, in this theory, rotations are left with no essential role. Most formulations until recently have followed that direction. However, with the growing need to develop size-dependent mechanics theory, there comes an opportunity not only to advance the discipline, but also to reconnect with some fundamental notions of mechanics. We believe that, if possible, the four foundational quantities (i.e., force, displacement, couple, rotation) should be at the very heart of such a theory and that individual terms in virtual work, as well as the essential and natural boundary conditions, should have a clear physical meaning. Therefore, consistent continuum mechanics must align seamlessly with rigid body mechanics.

Beyond this, there should always be an inner beauty and natural simplicity to mechanics, which is what attracts many of us to this field. The formulations presented in Neff et al. (2015a), and in the other papers in their recent series, cannot possibly point toward the future of mechanics. For example, the boundary conditions defined in equations (4) and (5) of Neff et al. (2015b) are far too complicated and non-physical. Moreover, one cannot hope to prove a consistent theory wrong by patching together several inconsistent theories, as those authors have attempted. There must instead be simple, elegant explanations of size-dependent response that will lead to a meaningful, self-consistent description of continua at the finest scales. Furthermore, we should note that the development of Neff et al. (2015a) is limited to linear isotropic elasticity, rather than providing generality for continuum mechanics as a whole. In this paper, we will not dwell on the details of Neff et al. (2015a), but instead focus on presenting consistent couple stress theory (Hadjesfandiari and Dargush, 2011), as clearly and concisely as possible. However, we also will examine the inconsistent deviatoric symmetric couple stress theory in this paper, as this helps to clarify the required consistency in a continuum mechanics theory.

It should be noted that elements of the consistent couple stress theory are based on the work of Mindlin and Tiersten (1962) and Koiter (1964), which use the four foundational continuum mechanical quantities (i.e., force, displacement, couple, rotation), without recourse to any additional degrees of freedom. This means the Mindlin-Tiersten-Koiter theory is based implicity on the rigid body portion of motion of infinitesimal elements of matter at each point of the continuum. In these important developments, Mindlin, Tiersten and Koiter correctly established that five geometrical and five mechanical boundary conditions can be specified on a smooth surface. However, their final theory suffers from some serious inconsistencies and difficulties with the underlying formulations, which may be summarized as follows:

1. The presence of the body couple in the relation for the force-stress tensor in the original theory[1];

2. The indeterminacy in the spherical part of the couple-stress tensor;

3. The inconsistency in boundary conditions, since the normal component of the couple-traction vector appears in the formulation.

This inconsistent theory is called the indeterminate couple stress theory in the literature (Eringen, 1968). Remarkably, consistent couple stress theory resolves all three of these inconsistencies with fundamental consequences. We notice that the major triumph in this development is discovering the skew-symmetric character of the couple-stress tensor. The important step in this discovery is to invoke the fundamental continuum mechanics hypothesis that the theory must be valid not only for the actual domain, but in all arbitrary subdomains. (This, of course, is exactly the same hypothesis that allows us to pass from global balance laws to the usual local differential forms.) Our involvement with boundary integral equations and the passion of the first author with the concept of rotation throughout mechanics and physics (Hadjesfandiari, 2013) have provided the necessary background. Furthermore, we should note that consistent couple stress theory offers a fundamental basis for the development of size-dependent theories in many multi-physics disciplines that may govern the behavior of continua at the smallest scales.

---

[1] In our previous work on couple stress theory, we incorrectly stated that the body-couple appeared in the *constitutive* relation for the force-stress tensor in the Mindlin-Tiersten-Koiter theory. We thank the authors of Neff et al. (2015a) for pointing out this error.

The paper is organized as follows. In Section 2, we consider consistent couple stress theory in detail and clarify some apparent ambiguities left in the original presentation. In this section, we demonstrate that a consistent continuum mechanics theory should be based on the rigid body portion of motion for infinitesimal elements of matter at each point in the continuum and the relative displacement and rotation of these elements at adjacent points. Then, we establish the skew-symmetric character of the couple-stress tensor based on the requirements for having consistent well-posed boundary conditions. After that in Section 3, we examine the deviatoric symmetric couple stress theory, which helps us to understand some inconsistencies that have plagued different size-dependent continuum mechanics theories. Finally, Section 4 contains a summary and some general conclusions.

## 2. Consistent couple stress theory

Consider a material continuum occupying a volume $V$ bounded by a surface $S$ with outer unit normal $n_i$, as shown in Fig. 1, under the influence of external loading, such as surface tractions and body-forces. Let us begin with the governing partial differential equations for couple stress theory representing the force and moment balance equations under quasistatic conditions, which can be written, respectively, as:

$$\sigma_{ji,j} + \overline{F}_i = 0 \tag{1}$$

$$\mu_{ji,j} + \varepsilon_{ijk}\sigma_{jk} = 0 \tag{2}$$

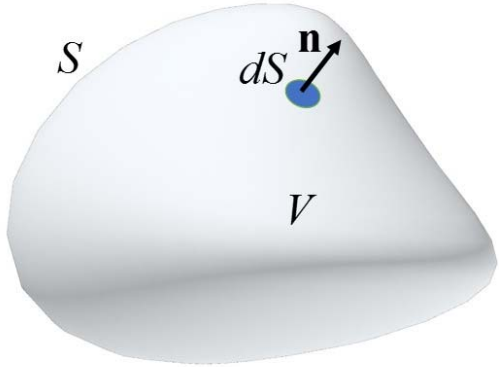

**Fig. 1.** The body configuration.

where $\sigma_{ij}$ represents the true (polar) force-stress tensor, $\mu_{ij}$ is the pseudo (axial) couple-stress tensor, $\bar{F}_i$ is the specified body-force density and $\varepsilon_{ijk}$ is the Levi-Civita alternating symbol. Any specified body-couple density can be rewritten in terms of body-force density and tangential force-tractions on the surface, and so does not appear explicitly in the governing equations. Here and throughout the remainder of this paper standard indicial notation is used with summation over repeated indices and with indices appearing after a comma representing spatial derivatives. Please note that there is no need to complicate the presentation with concepts from Lie algebra, orthogonal Cartan decompositions or generalized coordinates. These are completely superfluous to the important arguments and only tend to distract.

We notice that the force and moment balance laws (1) and (2) are the governing equations for translational and rotational equilibrium developed by considering infinitesimal elements of matter. Therefore, we are concerned with the rigid body portion of motion of infinitesimal elements of matter at each point of the continuum. However, the force and moment balance laws (1) and (2) do not by themselves have a unique solution for distribution of stresses in the continuum. For this purpose, we need to consider the deformation in terms of relative rigid body motion of infinitesimal elements of matter in the continuum under the influence of internal stresses. This provides us with the constitutive equations, which complete the set of equations to permit a unique solution of a well-posed boundary value problem. We consider next kinematics of a continuum.

In a consistent continuum representation, it is assumed that matter is continuously distributed in space, which requires the deformation to be specified completely by the continuous displacement field $u_i$. As a result, all kinematical quantities and measures of deformation must be derived from this displacement field. Fig. 2 allows us to visualize kinematics in the three-dimensional case. At each point, we define a rigid triad, which can be used to represent the rigid body portion of motion associated with infinitesimal elements at each point of the continuum. These rigid triads translate and rotate with the medium to provide the underlying rigid body portion of motion of each infinitesimal element, defined by the true (polar) displacement vector $u_i$ and the pseudo (axial) rotation vector $\omega_i$. Thus, the rigid body portion of motion of infinitesimal elements of matter at each point in three-dimensional space is described by six degrees of freedom, involving three translational $u_i$ and three rotational $\omega_i$ degrees of freedom. However, the continuity of matter within a continuum description restrains the rotation $\omega_i$ to equal one-half the curl of the displacement, which of course shows that the rotation field $\omega_i$ is not independent of the displacement field $u_i$. This latter aspect was missed by Cosserat and Cosserat (1909) and by those advocating for micropolar and related theories. Nonetheless, the Cosserats should be credited with the concept of the rigid triad and in elevating the role of angular momentum balance in continuum mechanics.

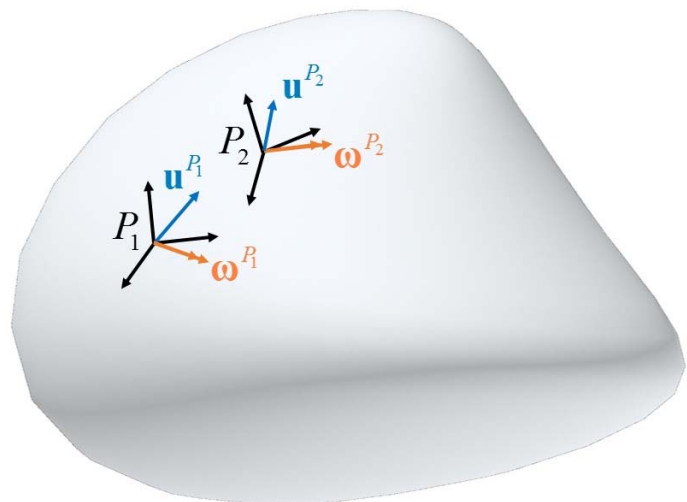

**Fig. 2.** The kinematics of a continuum.

These arguments indicate that rigid body motion is so fundamental in understanding continuum mechanics that the quantities $u_i$ and $\omega_i$ must directly appear as primary variables. Furthermore, this all suggests that consistent continuum mechanics theory should be developed as an extension of rigid body mechanics, which then is recovered in the absence of deformation.

To complete the deformation analysis, we need to define suitable measures or metrics of deformation based on the relative rigid body motion of triads at adjacent points of the continuum. For this purpose, consider two infinitesimal elements of matter at arbitrary points $P_1$ and $P_2$, as shown in Fig. 2. The displacements and rotations of these elements (or triads) are denoted by $u_i^{P_1}$ and $\omega_i^{P_1}$ at point $P_1$, and $u_i^{P_2}$ and $\omega_i^{P_2}$ at point $P_2$. Therefore, the relative translation $\Delta u_i$ and rotation $\Delta \omega_i$ of the element $P_2$ relative to the element $P_1$ can be expressed as

$$\Delta u_i = u_i^{P_2} - u_i^{P_1} = \int_{P_1}^{P_2} u_{i,j} dx_j \tag{3}$$

and

$$\Delta \omega_i = \omega_i^{P_2} - \omega_i^{P_1} = \int_{P_1}^{P_2} \omega_{i,j} dx_j \tag{4}$$

respectively. These equations show that the relative rigid body motion of infinitesimal elements of matter is described by the gradient of the translation tensor $u_{i,j}$ and the gradient of the rotation tensor $\omega_{i,j}$. This result suggests that the tensors $u_{i,j}$ and $\omega_{i,j}$ are of prime importance in deformation analysis and should appear in defining the measures of deformation. It should be mentioned that in some time-dependent phenomena, such as viscoelasticity and fluid mechanics, this relative motion is described instead by the velocity and angular velocity or vorticity vectors.

We recall that in classical continuum mechanics, the symmetric part of $u_{i,j}$, the strain tensor $e_{ij}$, accounts for the deformation by measuring stretch of straight element lines. This means we only consider the translating relative motion from (3) of infinitesimal elements of matter within the continuum in the classical theory. On the other hand, in size-dependent continuum mechanics, we also need to consider the relative rotation of infinitesimal elements (i.e., the relative rotation of the rigid triads), as in (4). This necessitates the contribution of the gradient of rotation tensor $\omega_{i,j}$ in the definition of the bending metric or measure of deformation, which ultimately will reduce to curvatures, as we shall see. From this kinematical analysis, other gradients of deformations, such as $e_{ij,k}$ and $\omega_{i,jk}$, do not appear as measures of deformation in a consistent continuum mechanics.

Although the gradient of deformation tensor $u_{i,j}$ is important in the analysis of deformation, even in the classical case, it is not in itself a suitable measure of deformation. In small deformation theory, this tensor can be decomposed into the true (polar) symmetric strain tensor $e_{ij}$ and the true (polar) skew-symmetric rotation tensor $\omega_{ij}$, where

$$e_{ij} = u_{(i,j)} = \frac{1}{2}\left(u_{i,j} + u_{j,i}\right) \tag{5}$$

$$\omega_{ij} = u_{[i,j]} = \frac{1}{2}\left(u_{i,j} - u_{j,i}\right) \tag{6}$$

Notice that parentheses around a pair of indices denote the symmetric part of the second order tensor, whereas square brackets indicate the skew-symmetric part. Then, the pseudo (axial) rotation vector $\omega_i$ discussed above, dual to the true skew-symmetric rotation tensor $\omega_{ij}$, is defined as

$$\omega_i = \frac{1}{2}\varepsilon_{ijk}\omega_{kj} = \frac{1}{2}\varepsilon_{ijk}u_{k,j} \tag{7}$$

where we also have the relation

$$\omega_{ji} = \varepsilon_{ijk}\omega_k \tag{8}$$

Now the principle of virtual work can be developed by first multiplying (1) and (2) by energy conjugate virtual quantities and then integrating over the volume $V$. In the case of couple stress theory, these energy conjugates must be the true (polar) virtual displacement $\delta u_i$ and the pseudo (axial) virtual rotation $\delta\omega_i$ for equations (1) and (2), respectively. Here we should note that (1) is a true (polar) vector equation, while (2) is in the form of a pseudo (axial) vector relation.

Multiplication by the conjugate virtual fields defined above produces in both cases a true scalar, which represents a virtual work density that is then integrated over the domain.

In this manner, the development of the principle of virtual work begins by writing:

$$\int_V \left[ \left( \sigma_{ji,j} + \bar{F}_i \right) \delta u_i + \left( \mu_{ji,j} + \varepsilon_{ijk} \sigma_{jk} \right) \delta \omega_i \right] dV = 0 \tag{9}$$

Note that this approach will provide a formulation with the corresponding real kinematic fields as the essential variables. Thus, the displacements and rotations will become the primary degrees of freedom and we will have a continuum formulation based upon the fundamental entities of mechanics.

For suitably differentiable fields, we may rewrite (9) by introducing the relations

$$\sigma_{ji,j} \, \delta u_i = \left( \sigma_{ji} \, \delta u_i \right)_{,j} - \sigma_{ji} \delta u_{i,j} \tag{10}$$

$$\mu_{ji,j} \, \delta \omega_i = \left( \mu_{ji} \, \delta \omega_i \right)_{,j} - \mu_{ji} \, \delta \omega_{i,j} \tag{11}$$

which after invoking the divergence theorem provides the following:

$$\int_S \left[ t_i^{(n)} \, \delta u_i + m_i^{(n)} \, \delta \omega_i \right] dS + \int_V \left[ -\sigma_{ji} \, \delta u_{i,j} + \bar{F}_i \, \delta u_i - \mu_{ji} \, \delta \omega_{i,j} + \varepsilon_{ijk} \sigma_{jk} \, \delta \omega_i \right] dV = 0 \tag{12}$$

where the force-traction true (polar) vector and couple-traction pseudo (axial) vector are defined as

$$t_i^{(n)} = \sigma_{ji} \, n_j \tag{13}$$

$$m_i^{(n)} = \mu_{ji} \, n_j \tag{14}$$

respectively, with $n_j$ representing the unit outward normal to the surface $S$. Fig. 3 illustrates force-traction and couple-traction vectors at an arbitrary location on the surface.

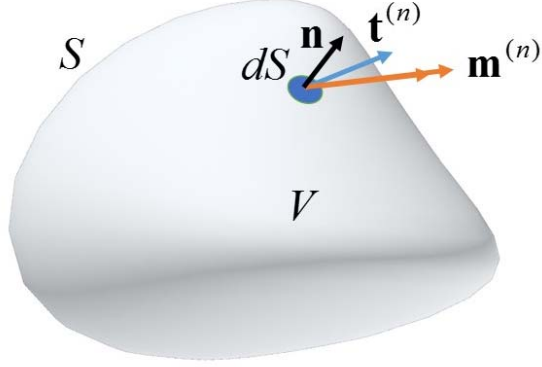

**Fig. 3.** Arbitrary force-traction and couple-traction vectors on surface.

However, $\delta \omega_{ij}$ is dual to the axial vector $\delta \omega_k$, such that

$$\delta \omega_{ij} = \varepsilon_{jik} \delta \omega_k \tag{15}$$

Then, (12) reduces to the following:

$$\int_V \left[ \sigma_{(ji)} \delta e_{ij} + \mu_{ji} \delta \omega_{i,j} \right] dV = \int_S \left[ t_i^{(n)} \delta u_i + m_i^{(n)} \delta \omega_i \right] dS + \int_V \left[ \bar{F}_i \delta u_i \right] dV \tag{16}$$

If one places a restriction now to kinematically compatible virtual fields on the boundary, then (16) would represent the principle of virtual work from the Mindlin and Tiersten (1962) indeterminate couple stress theory. We notice that the left hand side of (16) shows that the strain tensor $e_{ij}$ is energetically conjugate to the symmetric part of force-stress tensor $\sigma_{(ji)}$, which is consistent with our notion in classical continuum mechanics. Additionally, this relation shows that $\mu_{ji}$ and $\omega_{i,j}$ are energy conjugate tensors. This confirms our prediction that $\omega_{i,j}$ should contribute in the definition of the bending measure of deformation. Mindlin and Tiersten (1962), and Koiter (1964) considered the deviatoric tensor $\omega_{i,j}$ as the bending measure of deformation. However, this creates some inconsistencies in the formulation, such as the indeterminacy in the spherical part of the couple-stress tensor. Most importantly, the virtual work principle (16) shows that there is no room for strain gradients as fundamental measures of deformation in a consistent couple stress theory, as was concluded above in our kinematical analysis.

The right hand side of the virtual work principle (16) shows that the boundary conditions on the surface of the body can be either vectors $u_i$ and $\omega_i$ as essential (geometrical) boundary conditions, or $t_i^{(n)}$ and $m_i^{(n)}$ as natural (mechanical) boundary conditions. This apparently makes a total number of six boundary values for either case. However, this is in contrast to the number of independent geometric boundary conditions that can be imposed (Mindlin and Tiersten, 1962, Koiter, 1964). In particular, if components of $u_i$ are specified on the boundary surface, then the normal component of the rotation $\omega_i$ corresponding to twisting

$$\omega_i^{(n)} = \omega^{(nn)} n_i = \omega_k n_k n_i \tag{17}$$

where

$$\omega^{(nn)} = \omega_k n_k \tag{18}$$

cannot be prescribed independently. Therefore, the normal component $\omega^{(nn)}$ is not an independent degree of freedom, no matter whether the displacement vector $u_i$ is specified or not. However, the tangential component of rotation $\omega_i$ corresponding to bending, that is,

$$\omega_i^{(ns)} = \omega_i - \omega_i^{(n)} = \omega_i - \omega_k n_k n_i \tag{19}$$

represents two independent degrees of freedom in the global coordinate system, and may be specified in addition to $u_i$. As a result, the total number of geometric or essential boundary conditions that can be specified on a smooth surface is five.

Next, we let $m_i^{(nn)}$ and $m_i^{(ns)}$ represent the normal and tangential components of the surface couple-traction vector $m_i^{(n)}$, respectively. The normal component

$$m_i^{(nn)} = m^{(nn)} n_i \tag{20}$$

where

$$m^{(nn)} = m_k^{(n)} n_k = \mu_{ji} n_i n_j \tag{21}$$

causes twisting, while

$$m_i^{(ns)} = m_i^{(n)} - m^{(nn)} n_i = \mu_{kj} n_k \left( \delta_{ij} - n_j n_i \right) \tag{22}$$

is responsible for bending. Therefore, the boundary couple-traction virtual work in (16) can be written as

$$\begin{aligned}\int_S m_i^{(n)} \delta \omega_i dS &= \int_S m_i^{(nn)} \delta \omega_i^{(n)} dS + \int_S m_i^{(ns)} \delta \omega_i^{(ns)} dS \\ &= \int_S m^{(nn)} \delta \omega^{(nn)} dS + \int_S m_i^{(ns)} \delta \omega_i^{(ns)} dS\end{aligned} \tag{23}$$

As we know from theoretical mechanics, the generalized forces are associated only with independent generalized degrees of freedom, thus forming energetically dual or conjugate pairs. From the kinematic discussion above, $\omega^{(nn)}$ is not an independent generalized degree of freedom. Consequently, its corresponding generalized force $m^{(nn)}$ (i.e., the torsional component of the couple-traction) must be zero, that is

$$m^{(nn)} = m_k^{(n)} n_k = \mu_{ji} n_i n_j = 0 \quad \text{on } S \tag{24}$$

As a result, the boundary moment surface virtual work in (23) becomes

$$\int_S m_i^{(n)} \delta \omega_i dS = \int_S m_i^{(ns)} \delta \omega_i dS = \int_S m_i^{(ns)} \delta \omega_i^{(ns)} dS \tag{25}$$

This shows that a material in couple stress theory does not support independent distributions of normal surface twisting couple-traction $m^{(nn)}$, and the number of mechanical boundary conditions also is five. Consequently, while the force-traction may be in an arbitrary direction, the couple-traction must lie in the tangent plane, as shown in Fig. 4. This means a consistent couple stress theory must satisfy the boundary condition (24) automatically in its formulation.

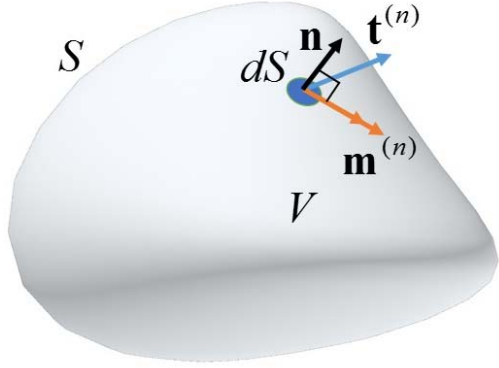

**Fig. 4.** Force-traction and tangential couple-traction vectors on surface.

This fundamental result was first established by Mindlin and Tiersten (1962) and more fully by Koiter (1964). However, the non-symmetric form of the couple-stress tensor $\mu_{ij}$ in their theory does not satisfy this requirement directly in the formulation, where a generally non-zero distribution of $m^{(nn)}$ seemingly can be applied on the boundary surface $S$. In fact, the fundamental implication of (24) as a constraint on the form of $\mu_{ij}$ was not understood fully until recently.

To resolve this problem, Koiter (1964) proposed that a distribution of normal surface twisting couple-traction $m^{(nn)}$ on the actual surface $S$ be replaced by an equivalent shear stress distribution and a line force system. This is analogous to the transformation of twisting shear distribution to an equivalent vertical transverse shear force and end corner concentrated forces in Kirchhoff bending theory of plates. However, we notice that there is a fundamental difference between couple stress theory and the Kirchhoff bending theory of plates. The Kirchhoff plate theory is a structural mechanics approximation to a continuum mechanics theory obtained by enforcing a constrained deformation. Therefore, results from this plate theory are not valid on and around the boundary surface, and near concentrated point and line loads. It is a fact that the plate theory usually gives better results in the internal bulk of the plate far enough from boundary and concentrated loads. On the other hand, couple stress theory is a continuum mechanics theory itself and should be valid everywhere, including near to and on the boundary, without any approximation. After all, we expect that the size-dependency and effect of couple stresses are more important near boundary surfaces, holes and cracks. Therefore, a consistent couple stress continuum theory should treat all parts of a material body with the same mathematical rigor and should not be considered as a structural mechanics formulation. Nevertheless, this fundamental difficulty with boundary condition (24) and its impact on the formulation was not appreciated at the time. It turns out that satisfying the condition (24) in a systematic way yields the consistent couple stress theory by revealing the fundamental character of the couple-stress pseudo tensor as follows.

We notice that by the fundamental continuum mechanics hypothesis, the principle of virtual work and its consequences are valid not only for the actual domain $V$, but for any arbitrary subdomain with volume $V_a$ having surface $S_a$, as shown in Fig. 5. Therefore, the normal surface twisting couple-traction $m^{(nn)}$ on the artificial surface $S_a$ must vanish, that is

$$m^{(nn)} = \mu_{ji} n_i n_j = 0 \quad \text{on} \ S_a \tag{26}$$

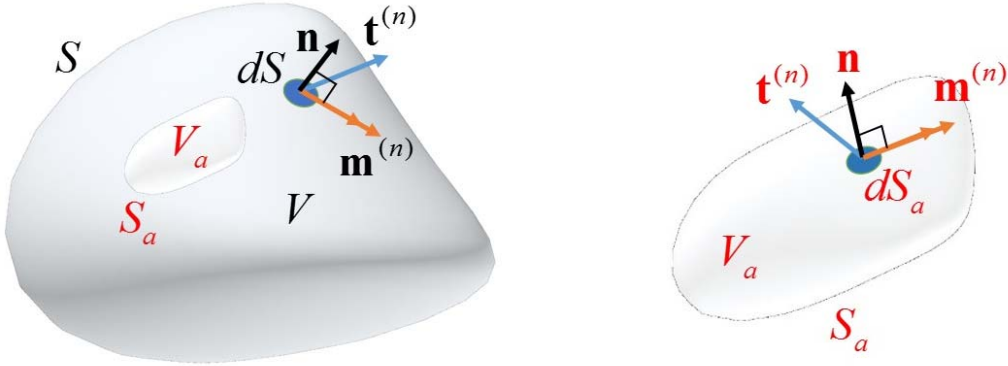

**Fig. 5.** The state of couple-traction $\mathbf{m}^{(n)}$ inside the body.

In the original Mindlin-Tiersten-Koiter theory, a generally non-zero distribution of $m^{(nn)}$ appears on the boundary surface $S_a$. However, we notice that the Koiter loading transformation method for this possible distribution of $m^{(nn)}$ on the artificial surface $S_a$ is incompatible with the arbitrariness of the surface $S_a$. This means that the couple stress distribution in the domain has to satisfy the condition (26) directly without recourse to any loading transformation. Thus, for any point on the arbitrary surface $S_a$ with unit normal $n_i$, $m^{(nn)}$ must vanish. This requires

$$m^{(nn)} = \mu_{ji} n_i n_j = 0 \quad \text{in} \quad V \tag{27}$$

However, in this relation, $n_i$ is arbitrary at each point; we may construct subdomains with any surface normal orientation at a point. Consequently, in (27), $n_i n_j$ is an arbitrary symmetric second order tensor of rank one at each point. Therefore, for (27) to hold in general, the couple stress pseudo tensor $\mu_{ij}$ must be skew-symmetric, that is

$$\mu_{ji} = -\mu_{ij} \tag{28}$$

This is the fundamental discovery of consistent couple stress theory, which shows that the couple-traction vector $m_i^{(n)}$ in (14) is tangent to the surface, thus creating purely a bending effect. We should emphasize that there is no mention of constitutive relations in any of this development, so that these results are in no way limited to linear elastic materials or to isotropic response. In this development, there are no additional assumptions beyond that of the continuum as a domain-based concept having no special characteristics associated with the actual bounding surface over any arbitrary internal surface.

The skew-symmetric character immediately resolves the indeterminacy problem. Since the diagonal components of the couple-stress tensor vanish, we notice that the couple-stress tensor automatically is determinate in this consistent couple-stress theory. Interestingly, this result indicates that there is an interrelationship between the consistent mechanical boundary condition (24) and the determinacy of the couple-stress tensor; resolving one, resolves the other. This is the amazing result of the fundamental hypothesis of continuum mechanics that the theory must be valid not only for the actual domain, but in all arbitrary subdomains. This realization is what was missed by Mindlin, Tiersten and Koiter in their quest for a consistent couple stress theory.

The components of the force-stress $\sigma_{ij}$ and couple-stress $\mu_{ij}$ tensors in this consistent theory are shown in Fig. 6. Since $\mu_{ij}$ is skew-symmetric, the couple-traction $m_i^{(n)}$ given by (14) is tangent to the surface. As a result, the couple-stress tensor $\mu_{ij}$ creates only bending couple-tractions on any arbitrary surface. The force-traction $t_i^{(n)}$ and the consistent bending couple-traction $m_i^{(n)}$ acting on an arbitrary surface with unit normal vector $n_i$ are shown again in Fig. 7.

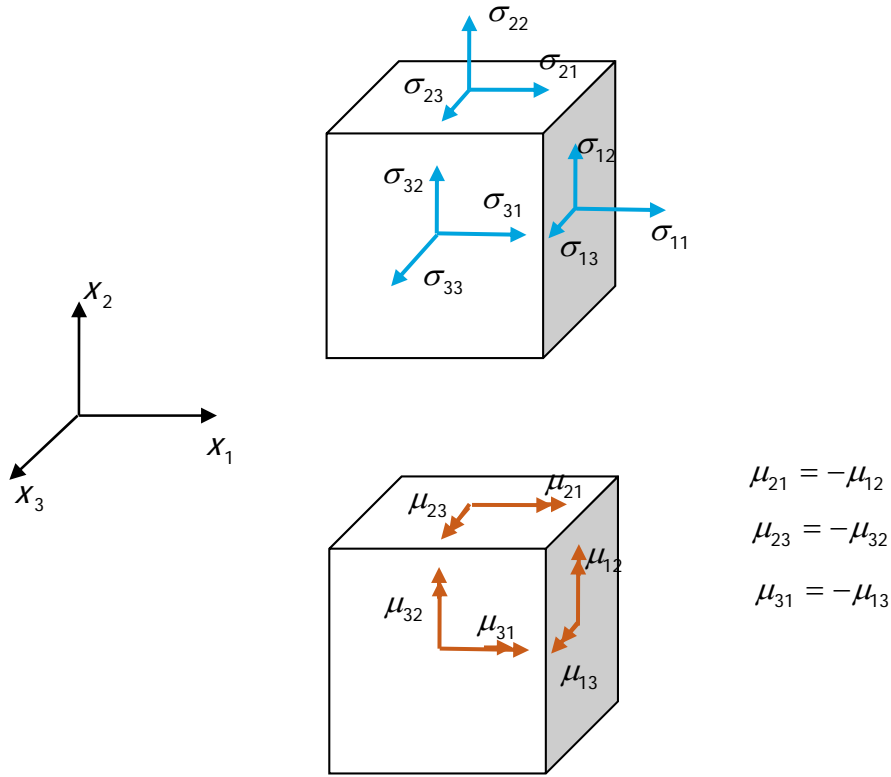

**Fig. 6.** Components of force- and couple-stress tensors in consistent couple stress theory.

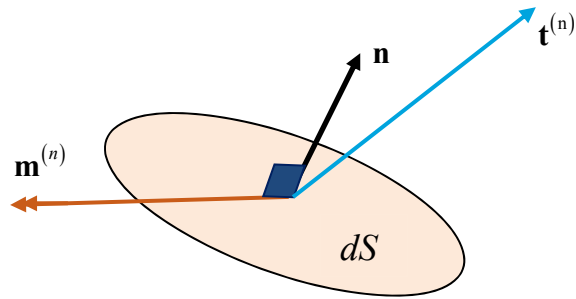

**Fig. 7.** Force-traction $\mathbf{t}^{(n)}$ and the consistent bending couple-traction $\mathbf{m}^{(n)}$.

It should be noticed that in this consistent continuum theory, the shear force-stresses, i.e. the tangential components of $t_i^{(n)}$ on any surface, completely account for the torsional loading in the material, a character similar to classical continuum mechanics.

The true (polar) couple-stress vector $\mu_i$ dual to the pseudo-tensor $\mu_{ij}$ is defined as

$$\mu_i = \frac{1}{2}\varepsilon_{ijk}\mu_{kj} \tag{29}$$

where we also have the relation

$$\mu_{ji} = \varepsilon_{ijk}\mu_k \tag{30}$$

Consequently, the surface couple-traction vector tangent to the surface $m_i^{(n)}$ reduces to

$$m_i^{(n)} = m_i^{(ns)} = \mu_{ji}n_j = \varepsilon_{ijk}n_j\mu_k \tag{31}$$

Here, it should be emphasized that the couple-traction vector $m_i^{(n)}$ is a pseudo vector, whereas the couple-stress vector $\mu_i$ is a true vector.

Since the couple-stress tensor is skew-symmetric, we can obtain the skew-symmetric part of the force-stress tensor from (2), as

$$\sigma_{[ji]} = \frac{1}{2}\varepsilon_{ijk}\mu_{lk,l} = -\mu_{[i,j]} \tag{32}$$

Thus, for the total force-stress tensor, we have

$$\sigma_{ji} = \sigma_{(ji)} + \frac{1}{2}\varepsilon_{ijk}\mu_{lk,l} = \sigma_{(ji)} - \mu_{[i,j]} \tag{33}$$

Therefore, there are nine independent stress components in consistent couple stress theory or general size-dependent continuum mechanics. This includes six components of $\sigma_{(ji)}$ and three components of $\mu_i$.

Interestingly, the relation (32) can be elaborated further if we consider the pseudo (axial) vector $s_i$ dual to the skew-symmetric part of the force-stress tensor $\sigma_{[ij]}$, where

$$s_i = \frac{1}{2}\varepsilon_{ijk}\sigma_{[kj]} \tag{34}$$

Then, by using (32) in (34), we obtain

$$s_i = \frac{1}{2}\varepsilon_{ijk}\mu_{k,j} \tag{35}$$

It is amazing to notice that the apparently complicated moment equilibrium equation (2) reduces to the simple curl relation (35). This is the result of the skew-symmetric character of the couple-stress tensor.

Consequently, the linear equation of equilibrium reduces to

$$[\sigma_{(ji)} + \mu_{[j,i]}]_{,j} + \overline{F}_i = 0 \tag{36}$$

which shows that there are only three independent equilibrium equations. Therefore, we must obtain the necessary extra six remaining equations from constitutive relations.

Now by returning to the virtual work principle (16), we notice that the skew-symmetric part of the tensor $\omega_{i,j}$, namely,

$$\kappa_{ij} = \omega_{[i,j]} = \frac{1}{2}\left(\omega_{i,j} - \omega_{j,i}\right) \tag{37}$$

is the consistent curvature pseudo tensor. Further inspection shows that the pseudo tensor $\kappa_{ij}$ is the mean curvature tensor, which represents the pure bending of material (Hadjesfandiari and Dargush, 2011). Moreover, the true (polar) mean curvature vector $\kappa_i$ dual to the pseudo-tensor $\kappa_{ij}$ is defined as

$$\kappa_i = \frac{1}{2}\varepsilon_{ijk}\kappa_{kj} \tag{38}$$

where we also have the relation

$$\kappa_{ji} = \varepsilon_{ijk}\kappa_k \tag{39}$$

After some manipulation, (38) can be written as

$$\kappa_i = \frac{1}{2}\omega_{ji,j} = \frac{1}{4}\left(u_{j,ji} - \nabla^2 u_i\right) \tag{40}$$

Interestingly, the mean curvature vector also can be expressed in terms of strain gradients as

$$\kappa_i = \frac{1}{2}\left(e_{kk,i} - e_{ik,k}\right) \tag{41}$$

Here, we should emphasize that this relation shows the curvature vector $\kappa_i$ cannot be expressed in terms of the arbitrary gradients of strain $e_{ij,k}$, but rather a very specific set of derivatives.

On the other hand, we notice that the symmetric part of the tensor $\omega_{i,j}$, that is,

$$\chi_{ij} = \omega_{(i,j)} = \frac{1}{2}\left(\omega_{i,j} + \omega_{j,i}\right) \tag{42}$$

is the torsion pseudo tensor (Hadjesfandiari and Dargush, 2011). The skew-symmetric character of the couple-stress tensor necessitates that the symmetric torsion tensor $\chi_{ij}$ does not contribute as a fundamental measure of deformation in a consistent couple stress theory.

Now by assuming kinematically compatible virtual fields in (16), the principle of virtual work balancing internal and external contributions is written:

$$\delta W_{int} = \delta W_{ext} \tag{43}$$

$$\int_V \left[\sigma_{ji}\,\delta e_{ij} + \mu_{ji}\,\delta\kappa_{ij}\right]dV = \int_{S_t}\left[\bar{t}_i^{(n)}\,\delta u_i\right]dS + \int_{S_m}\left[\bar{m}_i^{(ns)}\,\delta\omega_i^{(ns)}\right]dS + \int_V\left[\bar{F}_i\,\delta u_i\right]dV \tag{44}$$

where $\bar{t}_i^{(n)}$ and $\bar{m}_i^{(ns)}$ represent the prescribed force-tractions on $S_t$ and tangential couple-tractions on $S_m$, respectively, while $\delta\omega_i^{(ns)}$ are the tangential components of virtual rotation. Note that since $\delta e_{ij}$ is symmetric, only the symmetric part of the force-stress tensor $\sigma_{ij}$ contributes in (44).

Interestingly, the following observations can be made from our development, which demonstrate the inner beauty and natural simplicity of consistent continuum mechanics:

1. In classical continuum mechanics, there are no couple-stresses, such that $\mu_{ij} = 0$. As a result, the force-stress tensor $\sigma_{ij}$ is symmetric.

2. In couple stress continuum mechanics, the force-stress tensor $\sigma_{ij}$ is not symmetric, whereas the couple-stress tensor $\mu_{ij}$ is skew-symmetric. In addition, the skew-symmetric part of force-stress tensor $\sigma_{ij}$ is expressed in terms of the couple-stress tensor $\mu_{ij}$ via the elegant curl relation (35).

This result shows that both classical and couple stress continuum mechanics enjoy some level of symmetry in their inner structures.

We have demonstrated that in consistent continuum mechanics, we must consider the rigid body portion of motion of infinitesimal elements of matter (or rigid triads) at each point of the continuum. Therefore, in this consistent couple stress theory, the displacements and rotations provide the primary degrees of freedom. This is entirely compatible with the fundamental kinematic variables in classical mechanics, which define directly all of the basic rigid body motion. We also notice that the number of basic conservation laws of linear (1) and angular (2) momentum at each point is consistent with those for a rigid body.

The essential boundary conditions on a smooth surface in this couple stress theory for three-dimensional problems become the three displacements and two tangential rotations to form a set of five independent quantities. Meanwhile, natural boundary conditions consist of the force-traction vector with three independent components and the tangential couple-traction vectors to apply bending. As mentioned previously, this result was actually established by Mindlin, Tiersten and Koiter. Unfortunately, they did not realize that satisfying these boundary conditions in a systematic manner reveals the determinate skew-symmetric nature of the couple-stress tensor. Instead, by considering a general non-symmetric character for the couple-stress tensor, Koiter approximately enforced the required boundary conditions by using the loading transformation method from structural mechanics. However, the resulting couple stress theory was indeterminate and inconsistent. We notice that in the classical continuum mechanics theory, we only consider the motion of points or the relative translational rigid body portion of motion of infinitesimal elements of the continuum. As a result, the rotations are left with no essential role and the displacements become the primary degrees of freedom in this theory.

What could be more beautifully-consistent and physically-motivating for the definition of continuum boundary value problems than to base the theory on the four central quantities of mechanics? These are exactly the quantities, which describe the rigid body portion of motion of infinitesimal elements of matter at each point of the continuum. Fundamental solutions, variational principles, boundary integral representations, finite element methods, boundary element methods, finite difference methods, and solutions to a significant number of boundary value problems already have been developed for this consistent couple stress theory, within the context of both solid and fluid mechanics. Additional work is underway, as are physical experiments, to assess critically these formulations. Time will tell to what extent this self-consistent theory aligns with nature.

## 3. Deviatoric symmetric couple stress theory

Perhaps we should emphasize a further point. In consistent couple stress theory, the diagonal components of the couple-stress tensor always vanish due to the skew-symmetric character. Consequently, the determinate couple-stress tensor is also deviatoric. Therefore, we may conclude that the deviatoric skew-symmetric couple stress theory is the fully consistent and determinate theory. This is in contrast to the deviatoric symmetric couple stress theory, which suffers from many inconsistences. We examine this theory in detail in the following, as this might be helpful in appreciating more deeply the consistency and beauty of skew-symmetric couple stress theory.

Neff et al. (2009) support a theory based on the deviatoric (trace free) symmetric couple-stress tensor. This theory is also related to the work of Yang et al. (2002), which is commonly called the modified couple stress theory. In their development, Yang et al. (2002) consider an extra equilibrium equation for the moment of couples, in addition to the two equilibrium equations of the classical continuum. Of course, this additional law has no support in physical reality.

However, application of this unsubstantiated equilibrium equation, apparently leads to a symmetric couple-stress tensor, that is

$$\mu_{ji} = \mu_{ij} \tag{45}$$

The main motivation for Yang et al. (2002) in their development has been to reduce the number of couple-stress material constants for linear isotropic elastic material from two in the original Mindlin-Tiersten-Koiter theory to only one constant. For this theory, the virtual work principle (16) shows that the symmetric tensor $\chi_{ij}$ is the corresponding curvature tensor in this theory. However, we notice that

$$\chi_{ii} = \omega_{i,i} = 0 \tag{46}$$

which shows that the tensor $\chi_{ij}$ is deviatoric, and thus is specified by only five independent components. As a consequence, all the inconsistencies in Mindlin-Tiersten-Koiter theory, such as the indeterminacy in the couple-stress tensor and the appearance of $m^{(nn)}$ on the bounding surface $S$, unfortunately remain intact in this theory. Although, Yang et al. (2002) do not offer any reason for the disappearance of the indeterminate spherical part of the couple-stress tensor, many proponents of this theory assume the couple-stress tensor is also deviatoric, that is,

$$\mu_{ii} = \mu_{11} + \mu_{22} + \mu_{33} = 0 \tag{47}$$

There have been some doubts about the validity of the fundamental aspects of the deviatoric symmetric couple stress theory. As mentioned, the symmetry character of the couple-stress tensor in this theory is the consequence of the peculiar equilibrium equation for the moment of couple, besides the two conventional force and moment balance laws. However, this requirement is an additional condition, which is not derived by any principle of classical mechanics, as mentioned by Lazopoulos (2009). This simply shows that modified couple stress theory is not consistent with basic rigid body mechanics. For more explanation about this fundamental inconsistency, see Hadjesfandiari and Dargush (2014). However, there are some other issues with this theory, which we examine next.

First, we notice that a theory based on the constrained deviatoric (trace free) symmetric couple-stress tensor cannot be physically acceptable. We demonstrate this by using physical contradiction. If we assume the couple-stress tensor $\mu_{ij}$ is deviatoric and symmetric, it can also be diagonalized by choosing the coordinate system $x_1 x_2 x_3$, such that the coordinate axes $x_1$, $x_2$ and $x_3$ are along its orthogonal eigenvectors or principal directions. Therefore, in this coordinate system, the couple-stress tensor $\mu_{ij}$ is represented by

$$[\mu_{ij}] = \begin{bmatrix} \mu_{11} & 0 & 0 \\ 0 & \mu_{22} & 0 \\ 0 & 0 & \mu_{33} \end{bmatrix} \quad (48)$$

where the diagonal components $\mu_{11}$, $\mu_{22}$ and $\mu_{33}$ are the torsional couple-stress components around the coordinate axes $x_1$, $x_2$ and $x_3$, respectively. However, from a practical view, we notice that the loading along these directions are independent. This means we are allowed to exert torsion couple-stress in any direction; its amount is arbitrary. Therefore, if we can exert torsional couple-stresses $\mu_{11}$, $\mu_{22}$ and $\mu_{33}$ on some element of the matter, these three components must be independent of each other. This physical fact contradicts the mathematical deviatoric condition expressed by (47). Therefore, couple stress theory with a deviatoric symmetric couple-stress tensor is inconsistent and cannot be accepted on physical grounds.

We also notice that the symmetric tensor $\chi_{ij}$ is the torsion pseudo-tensor representing the pure twist of material (Hadjesfandiari and Dargush, 2011). Since this tensor $\chi_{ij}$ is symmetric, it can also be diagonalized by choosing the coordinate system $x_1 x_2 x_3$, such that the coordinate axes $x_1$, $x_2$ and $x_3$ are along its orthogonal eigenvectors or principal directions. Therefore, in this coordinate system the torsion tensor $\chi_{ij}$ is represented by

$$[\chi_{ij}] = \begin{bmatrix} \chi_{11} & 0 & 0 \\ 0 & \chi_{22} & 0 \\ 0 & 0 & \chi_{33} \end{bmatrix} \quad (49)$$

where the diagonal components $\chi_{11}$, $\chi_{22}$ and $\chi_{33}$ are the torsions around the coordinate axes $x_1$, $x_2$ and $x_3$, respectively. Therefore, the torsion tensor (49) does not represent the bending deformation of the material at all. This fact also suggests that this tensor should not be chosen as the sole bending measure of deformation.

Therefore, the deviatoric symmetric or the modified couple stress theory not only inherits all inconsistences from indeterminate Mindlin-Tiersten-Koiter theory, but also suffers from new inconsistencies, which are summarized as follows:

1. The unsubstantiated additional artificial equilibrium of moment of couples in the set of fundamental equations;

2. The physical inconsistency of the constrained deviatoric symmetric couple-stress tensor $\mu_{ij}$;

3. The deviatoric symmetric torsion tensor $\chi_{ij}$ does not describe the bending deformation.

As a final issue, one might think that the indeterminacy of the spherical part of the couple-stress tensor is analogous to the behavior of an incompressible material under pressure. For an incompressible material, the incompressibility condition is

$$u_{i,i} = 0 \tag{50}$$

Assume the distribution of the constant pressure $p$, where

$$\sigma_{ij} = -p\delta_{ij} \tag{51}$$

Consequently, the normal force-traction on the surface is

$$t_i^{(n)} = -pn_i \tag{52}$$

We notice that the pressure stress distribution (51) does not contribute to the internal work, because we have for the internal compatible virtual work

$$\sigma_{ji}\delta e_{ij} = -p\delta u_{i,i} = 0 \tag{53}$$

As a result, this loading does not create any deformation in the body. However, we notice that an incompressible material is a mathematical concept, and physically does not exist. This means that the strain tensor $e_{ij}$ never becomes deviatoric in reality. The incompressibility condition (50) is just an artificial assumption to simplify cases of near-incompressibility. Interestingly, for the linear isotropic elastic materials, the incompressibility corresponds to Poisson ratio $\nu = \frac{1}{2}$, which is excluded based on energy considerations (Malvern, 1969).

On the other hand, we notice that the deviatoric character of the couple-stress tensor in Mindlin-Tiersten-Koiter and modified couple-stress theory is the direct result of deviatoric tensors $\omega_{ij}$ and $\chi_{ij}$, respectively, independent of the material behavior. It is this deviatoric character, which makes these tensors unsuitable as measures of bending deformation. Nevertheless, we have already established that the skew-symmetric mean curvature tensor $\kappa_{ij}$ is the consistent measure of bending deformation, which of course has no spherical part.

## 4. Conclusions

The recent papers by Neff et al. (2015a-c) have motivated us to reexamine continuum mechanics from a fundamental perspective. However, what is most fundamental in developing a continuum mechanics theory? Is it the definition of thermodynamic potentials? Balance laws? Boundary conditions? Virtual work? Of course, all of these are important, but we believe that first and foremost the development should be founded on concepts emanating from the classical mechanics of particles and rigid bodies, in which all variables have clear physical meaning. Thus, the fundamental objects of investigation in mechanics should be forces and couples, or their intensive continuum counterparts, namely, force-stresses and couple-stresses. Furthermore, the kinematic variables must be displacement and rotation, which are needed to describe rigid motion of entire bodies or, in the continuum case, of infinitesimal elements. Since the force and moment balance laws for these infinitesimal elements of matter are not sufficient to determine uniquely the distribution of stresses in the continuum, we need to consider deformation.

To gain a better understanding of the kinematics of deformation, we may envision a rigid triad associated with each infinitesimal element. However, the continuity of matter restrains the relative motion of these rigid triads, such that here, unlike in Cosserat theory, the triad translates and also rotates with each infinitesimal element. There is no independent rotation; rather the rotation of each infinitesimal element, and its attached rigid triad, is defined by one-half the curl of the displacement field. In classical continuum mechanics, the deformation then is attributed solely to the symmetric part of the relative translation of adjacent infinitesimal elements (or rigid triads). However, this is an incomplete picture, which assigns a minor ancillary role to rotations and indicates that classical Cauchy continuum mechanics is not fully aligned with particle and rigid body mechanics. We must extend this classical view to accommodate the relative rotation of these adjacent infinitesimal elements (or rigid triads) as well, and elevate rotations to the level of kinematic degrees of freedom, along with displacements. Thus, relative triad translation provides displacement gradients, which lead to the identification of strains, or stretches in principal directions, as the size-independent measure of deformation, exactly as in the classical theory. On the other hand, relative triad rotation offers rotation gradients as the candidate from which a size-dependent deformation measure can be derived.

Next, by giving careful consideration to the issue of independent boundary conditions on both the real surfaces and any arbitrary internal surface, we find that the normal twisting couple-traction must vanish on all surfaces. Satisfying this requirement in a systematic way restricts the form of the couple-stress tensor to be skew-symmetric. This is what was missed by Mindlin, Tiersten and Koiter in their quest for a consistent couple stress theory. Because of its skew-symmetric nature, the couple stress tensor also is automatically deviatoric without imposing non-physical constraints on the components. At once, this resolves all of the issues of inconsistency and indeterminacy that have plagued prior couple stress theories, including the original Mindlin-Tiersten-Koiter and modified couple stress theories. Furthermore, the deformation measure that is energy conjugate to the skew-symmetric couple stress tensor becomes the skew-symmetric part of the rotation gradient tensor, that is, the mean curvature tensor, which captures size-dependent bending deformation.

Finally, we may mention the interesting symmetries present in the two main continuum theories. In classical continuum mechanics, there are no couple-stresses, and the force-stress tensor is symmetric. On the other hand, in consistent couple stress continuum mechanics, the force-stress tensor is not symmetric, but the couple-stress tensor is skew-symmetric. This suggests once again that the mathematical description of nature may favor a certain level of symmetry and beauty in its inner structure.